\documentclass[aps,prl,reprint]{revtex4-1}
\usepackage{amsmath,amsfonts,amssymb}
\usepackage{blindtext}
\usepackage{graphicx,float}
\usepackage{epstopdf}
\usepackage[utf8]{inputenc}
\usepackage[encapsulated]{CJK}

\usepackage{color}

\newcommand{\kpr}{k^{\prime}}
\newcommand{\remove}[1]{}

\usepackage[normalem]{ulem} % for \sout

\begin{document}
%\title{Higher Order Correlations Distort Local Information in Networks}
\title{Neighbor-Neighbor Correlations Explain Measurement Bias in Networks}
\author{Xin-Zeng Wu$^{1,2}$} %(\begin{CJK}{UTF8}{bsmi}吳信增\end{CJK})
\author{Allon G.~Percus$^{3}$}
\author{Kristina Lerman$^{1}$}
%\email{ }
\affiliation{$^{1}$ Information Sciences Institute, University of Southern California, Marina del Rey, CA 90292}
\affiliation{$^{2}$ Department of Physics and Astronomy, University of Southern California, Los Angeles, CA 90089}
\affiliation{$^{3}$ Institute of Mathematical Sciences, Claremont Graduate University, Claremont, CA 91711}

\begin{abstract}
In numerous physical models on networks, dynamics are based on interactions that exclusively involve properties of a node's nearest neighbors.
%Individuals in complex networks often base their behavior on observations of their neighbors.  
However, a node's local view of its neighbors may systematically bias perceptions of network connectivity or the prevalence of certain traits. We investigate the \emph{strong friendship paradox}, which occurs when the majority of a node's neighbors have more neighbors than does the node itself.
%, and demonstrate its prevalence in real-world networks.
We develop a model to predict the magnitude of the paradox, showing that it is enhanced by negative correlations between degrees of neighboring nodes.  We then show that by including neighbor-neighbor correlations, which are degree correlations one step \emph{beyond} those of neighboring nodes, we accurately predict the impact of the strong friendship paradox in real-world networks. Understanding how the paradox biases local observations can inform better measurements of network structure and our understanding of collective phenomena.

\end{abstract}

\maketitle
Local interactions among nodes in a complex network can lead to an astounding array of collective phenomena.  Examples include viral outbreaks in social networks, cascading failures in the power grid and financial networks, synchronization of coupled oscillators, opinion dynamics and consensus formation in human groups. Researchers have linked the structure of complex networks to the dynamics of collective phenomena unfolding on them: highly connected nodes amplify viral outbreaks~\cite{Watts02,Kempe03,Lloyd-Smith05}, while %the network's
community structure affects the dynamics of synchronization~\cite{Lerman12pre} and the spread of social contagions~\cite{weng2013virality}.

A node's own local view of a network, however, may be systematically biased.
%different from the global ground truth.
%which may affect collective behaviors. also affects sampling
%Social scientists have identified
One source of bias is Feld's friendship paradox: the number of connections, or degree, of a node is smaller than the average of its neighbor's degrees~\cite{Feld91}.
Recently, more subtle forms of the paradox have been proposed.  The \emph{strong} friendship paradox~\cite{Kooti14icwsm} states that the degree of a node tends to be smaller than the \emph{median} of its neighbor's degrees. Roughly speaking, this is equivalent to the node having fewer neighbors than do a majority of its neighbors.  But unlike the original friendship paradox and some recent generalizations~\cite{Hodas13icwsm,Eom14,Jo14,Ross16},
%to attributes other than degree~\cite{Hodas13icwsm,Eom14,Jo14},
the strong friendship paradox does not arise as a straightforward result of sampling from skewed distributions~\cite{Kooti14icwsm}.
The strong friendship paradox can dramatically distort local %information 
measurements in a network, leading to the ``majority illusion''~\cite{Lerman16majority} in which a globally rare attribute may be overrepresented in the local neighborhoods of many nodes. %. As a consequence, many nodes will observe the majority of neighbors with the attribute, which can affect an individual node's interactions, as well as
%which may affect collective phenomena unfolding on the network.  
Physical systems whose dynamics are governed by majority rule---% have been studied in the literature, ranging
from Ising spin interactions~\cite{Krapivsky03} to more complex voting models~\cite{Liggett99}---may be affected by this paradox.

In this Letter, we develop a stochastic model to predict the magnitude of the strong friendship paradox.  Specifically, we show that: a) increasingly disassortative networks exhibit a larger paradox, and b) accurately modeling it requires considering degree correlations one step beyond those of neighboring nodes.

% quantify the strong friendship paradox and relate it to the network's structural properties.
% Globally, the fraction of nodes that experience the strong friendship paradox is:
Given a network with degree distribution $p(k)$, we define the global probability of the strong friendship paradox as
$P_{\mathrm{paradox}}=\sum_k p(k)f(k)$,
% $$
% P_{\mathrm{paradox}}=\sum_{k=k_{\mathrm{min}}}^{k_{\mathrm{max}}}p(k)f(k),
% $$
where
% Here, $p(k)$ is the degree distribution, i.e.,\ the probability to find a node with $k$ neighbors, and
$f(k)$ is the probability that a randomly chosen node with degree $k$ experiences the paradox. %, i.e., it %meaning that it %its degree is lower than the median degree of its neighbors:
% has a lower degree than the median degree of its neighbors:
Formally, we define
%\begin{align}
$$
\label{classsum}
f(k) \equiv P\bigl(\mathrm{Median}\{k^{\prime}_1,\cdots,k^{\prime}_k\}> k|k\bigr),
$$
where $k^{\prime}_i$ is the degree of the node's $i$th neighbor.
% \nonumber\\[1.1ex]
% &\approx P(k^{\prime}_{(\left\lceil\frac{k}{2}\right\rceil)}>k|k).
% \end{align}
% We used the notation of order statistics above, for example, ${\kpr}_{(1)}=\min({\kpr}_1,\cdots,{\kpr}_k)$, ${\kpr}_{(k)}=\max({\kpr}_1,\cdots,{\kpr}_k)$.

%To accurately estimate the strength of the paradox in a network requires knowing higher order degree correlations represented by $2K$ and $3K$ structure.
Of course, networks can have
% Networks often have \remove{higher order}
structure beyond that given by the degree distribution. The $dK$-series framework~\cite{Mahadevan06} specifies network structure as a series of joint degree distributions of subgraphs of $d$ nodes. Thus, a network's $1K$-structure is specified by the degree distribution $p(k)$. The $2K$-structure captures degree correlations of nodes in connected pairs.
%For example, a node may preferentially connect to other nodes with similar or very different degrees
%($2K$-structure), or have neighbors with similar or very different degrees ($3K$-structure).
This is specified by the joint degree distribution $e(k,{\kpr})$,
%% which is a matrix whose elements give
the probability that an edge links two nodes with degrees $k$ and ${\kpr}$.
It follows that the degree distribution of an edge's endpoint is
%% Neighbor degree distribution is a link of the two quantities mentioned above, defined as
$q(k)=\sum_{\kpr}e(k,k')=kp(k)/\langle k\rangle$.
% Degree correlations are measured globally by assortativity, $r_{kk'}$ which is positive in assortative networks where nodes of similar degree tend to link, and it is negative in disassortative networks, in which nodes with dissimilar degrees link.
% The assortativity dependence of the conventional friendship paradox has been studied mathematically and numerically ...~\cite{Jo14}.
%From here, people often define the neighbor distribution $q(k)$, which is the probability for a specific node to have a neighbor with degree $k$.
%\begin{equation}
%q(k)=\sum_{{\kpr}}e(k,{\kpr})\hspace{0.3in}P({\kpr}|k)=\frac{e(k,{\kpr})}{q(k)}
%\end{equation}
Similarly, a network's $3K$-structure is specified by the
% \remove{degree correlations of nodes in connected triplets, either chains or triangles}
joint degree distribution of connected triplets, either wedges or triangles.
We find that %such next-nearest neighbor degree
these higher-order degree correlations can be substantial in real-world networks, possibly reflecting their macroscopic organization into %communities
a core-periphery structure,
and that accounting for them is necessary for a quantitative understanding of the strong friendship paradox.
%In order to quantitatively understand the paradox, we need to account for these higher order degree correlations.

%The strong friendship paradox depends only on the comparison between the degrees of a node and its neighbors. For convenience, we define indicator functions $x_i$,  $i=1\cdots k$, to track the neighbors of a node of degree $k$:
%\begin{equation}
%x_i = \mathbf{1}_{k^{\prime}_i>k} = \left\{\begin{array}{cl} 1 & \mathrm{if}\;k^{\prime}_i>k\\ 0 & \mathrm{if}\;k^{\prime}_i\leq k \end{array}\right.
%\end{equation}
%This node is in paradox regime if $\langle x\rangle\equiv\frac{1}{k}\sum_{i=1}^k x_i>\frac{1}{2}$.

%\subsubsection{Binomial model}

The strong friendship paradox depends only on the comparison between the
degrees of a node and its neighbors.
The probability $Q_>$ that a node sees a neighbor with degree larger than its own can be written % using the indicator function
as:
\begin{align}
\label{eq:P(k_i>k)}
% \deleted{P(k^{\prime}_i>k)}
Q_> &= \sum_k \sum_{k^{\prime}>k}%P(k^{\prime}_i>k|k,k^{\prime}_i)
P(k^{\prime}|k)p(k) %\nonumber\\
%&= \langle k\rangle\sum_{k,k^{\prime}_i}\frac{\boldsymbol{1}_{k^{\prime}_i>k}}{k}e(k,k^{\prime}_i)
=\langle k\rangle\sum_k\sum_{k^{\prime}>k}\frac{e(k,k^{\prime})}{k},
% = \frac{\langle k\rangle}{2}\sum_{k,k^{\prime}_i\neq k}\frac{e(k,k^{\prime}_i)}{\min(k,k^{\prime}_i)}
\end{align}
since $P({\kpr}|k)={e(k,{\kpr})}/{q(k)}$.
This expression uses information about the network's $2K$-structure, %which captures degree correlations of connected pairs.
which is globally 
%Globally, such correlations are 
measured by the assortativity coefficient~\cite{newman2002}
$$
r=\frac{1}{\mathrm{Var}(k)}\sum_{k,{\kpr}}kk'\left[e(k,{\kpr})-q(k)q({\kpr})\right],
$$
where the variance of $k$ is taken with respect to the distribution $q(k)$.
% The assortativity dependence of the conventional friendship paradox has been studied \remove{mathematically and} numerically~\cite{Jo14}. \note{KL: Main conclusion from Jo14 study?}
In assortative networks ($r>0$), nodes preferentially link to other nodes with similar degree, while in disassortative networks ($r<0$), they prefer to link to others with dissimilar degree, e.g., high to low degree nodes.
Since $k$ is in the numerator of the sum for $r$ but in the denominator of
%the sum for $P(k^{\prime}_i>k)$,
Eq.~\ref{eq:P(k_i>k)},
given the normalization $\sum_{k,\kpr}e(k,{\kpr})=1$, we may expect
% since $r$ and $P(k^{\prime}_i>k)$ have opposite trends of weights on $k$,
disassorativity to magnify the paradox in networks, and
assortativity to suppress it.  Previous numerical results for
the conventional friendship paradox~\cite{Jo14} support this prediction. 

\paragraph*{2K Model.}
Given a randomly chosen node with degree $k$, define an indicator function $x_i$, $i=1\ldots k$, to track the degree of the node's $i$th neighbor:
\begin{equation}
\label{ind}
x_i = \mathbf{1}_{k^{\prime}_i>k} = \left\{\begin{array}{cl} 1 & \mathrm{if}\;k^{\prime}_i>k\\ 0 & \mathrm{if}\;k^{\prime}_i\leq k \end{array}\right.
\end{equation}
%The node is in the paradox regime if the indicator function averaged over all neighbors, $\overline{x}\equiv\frac{1}{k}\sum_{i=1}^k x_i$, satisfies $\overline{x}>\frac{1}{2}$.
To a close approximation (and exactly, for odd $k$), the node is in the paradox regime if $\overline{x}\equiv\frac{1}{k}\sum_{i=1}^k x_i>\frac{1}{2}$.

To understand how network structure affects the strong friendship
paradox, we now examine $\mu_x(k)$, the probability that a neighbor (say the $i$th one) of a randomly chosen degree-$k$ node has degree greater than $k$:
%\changed{given that there are $n(k)$ nodes in the network with the same degree}:
% When the degrees of neighbors can be considered as independent, identically distributed (i.i.d.) variables, this quantity can be written as:
%\changed{Equation Changed!!}
\begin{align}
\label{mu}
\mu_x(k) =
P(x_i=1 | k) =
P(k^{\prime}_i>k | k)
%&= \bar{\bar{x}}=\frac{1}{kn(k)}\sum_{\alpha=1}^{n(k)}\sum_{i=1}^kx_{\alpha i}\nonumber\\
%&=\frac{1}{2Eq(k)}\sum_{\alpha=1}^{n(k)}\sum_{i=1}^k\boldsymbol{1}_{k'_{\alpha i}>k}
=\sum_{k'>k}\frac{e(k,k')}{q(k)}
\end{align}

% Here, for convenience we defined an indicator function $x_i$,  $i=1\ldots k$ to track neighbors of a node of degree $k$:
% \begin{equation}
% x_i = \mathbf{1}_{k^{\prime}_i>k} = \left\{\begin{array}{cl} 1 & \mathrm{if}\;k^{\prime}_i>k\\ 0 & \mathrm{if}\;k^{\prime}_i\leq k \end{array}\right.
% \end{equation}
% A degree-$k$ node is in the paradox regime if the indicator function,
% averaged over all neighbors of a node, is $\overline{x}\equiv\frac{1}{k}\sum_{i=1}^k x_i>\frac{1}{2}$.
% effect of assorativity on strong friendship paradox
% Binomial model

If we assume that degrees of neighbors are %\deleted{i.i.d.}
independent and identically distributed random variables, the probability for a degree-$k$ node to observe the strong friendship paradox is then given by the binomial distribution:
\begin{align}
% f(k) &= P\left(\frac{1}{k}\sum_{i=1}^kx_i>\frac{1}{2}\right)=P\left(\sum_{i=1}^kx_i>\frac{1}{2}k\right)\nonumber\\
f(k) &= P\left(\overline{x}>\frac{1}{2}\right)=P\left(\sum_{i=1}^kx_i>\frac{k}{2}\right)\nonumber\\
&=\sum_{i=\lceil\frac{k+1}{2}\rceil}^k\binom{k}{i}\mu_x(k)^i[1-\mu_x(k)]^{k-i}.
\label{f(k)-2K}
\end{align}
For large $k$, $f(k)$ is close to Gaussian.
In terms of the normal distribution's cumulative distribution function $\Phi$,
\begin{equation}
f(k)=
1-\Phi\left\{\frac{1-2\mu_x(k)}{2\sqrt{\mu_x(k)[1-\mu_x(k)]}}\sqrt{k}\right\}.
\label{normalf}
\end{equation}
% \begin{align}
% \label{normalf}
% f(k)&=1-\Phi\left\{\frac{\frac{1}{2}-\mu_x(k)}{\sigma_x(k)}\right\}\\
% \label{normalf1}
% &=1-\Phi\left\{\frac{1-2\mu_x(k)}{2\sqrt{\mu_x(k)[1-\mu_x(k)]}}\sqrt{k}\right\}
% \end{align}

\begin{figure}
\centering
\begin{tabular}{cc}
\includegraphics[width=0.25\textwidth]{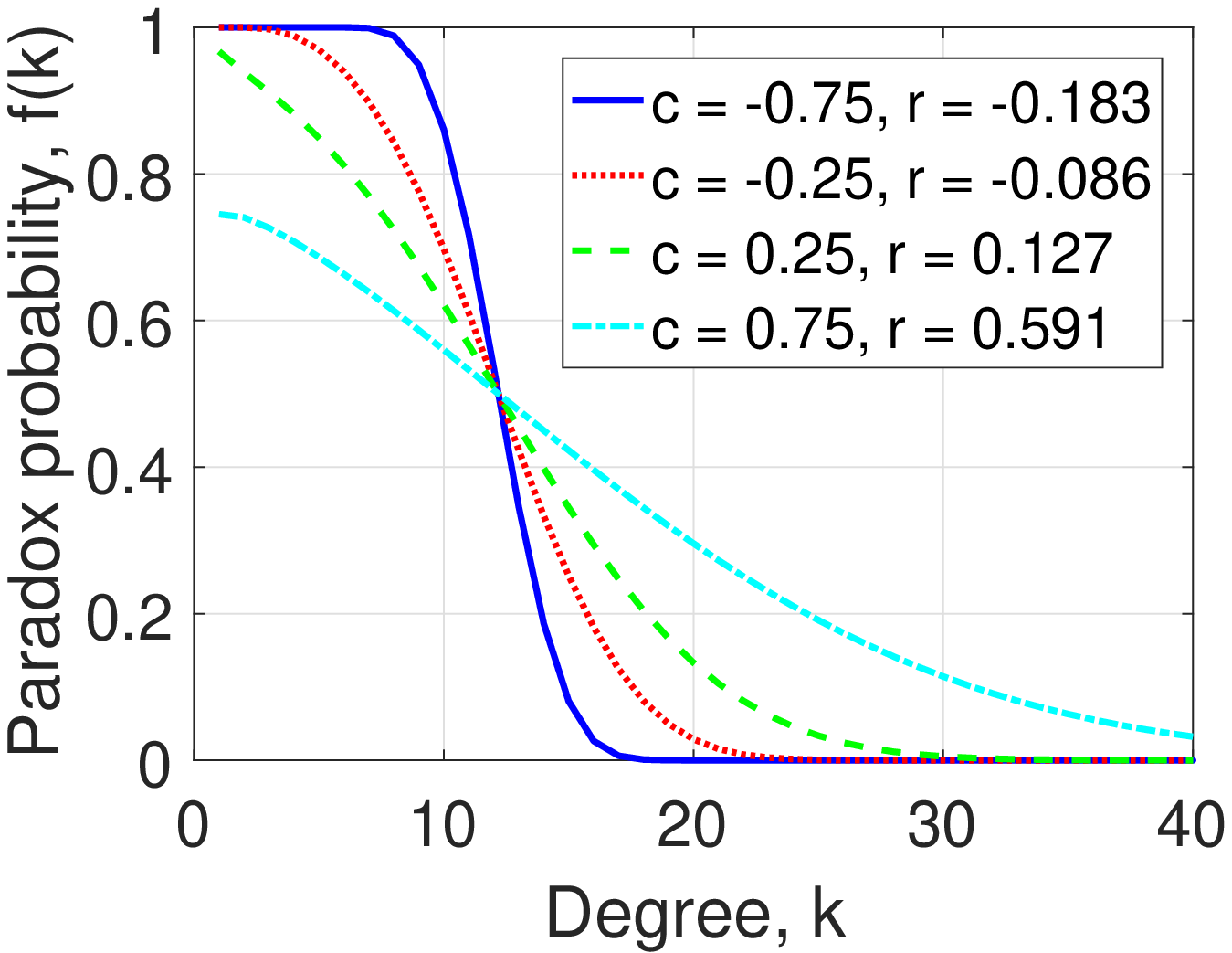}
&
\includegraphics[width=0.25\textwidth]{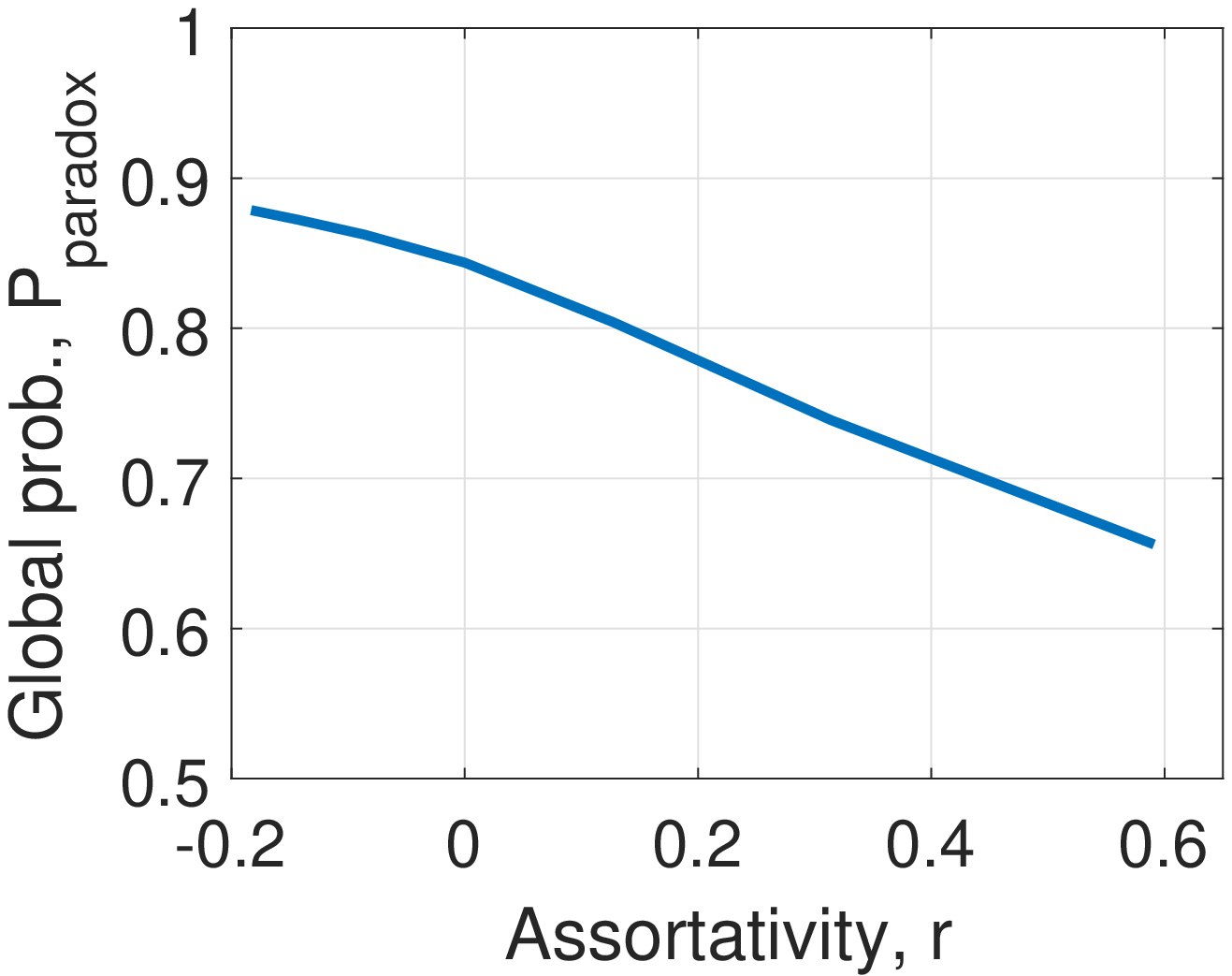}
\\
(a) & (b)
\end{tabular}
\caption{(Color online) Impact of assortativity on strong friendship paradox. The matrix $e(k,k')$ has bivariate log-normal distribution with parameters $m=2.5$, $s=1.25$, $c$ ranges from $-0.75$ to 0.75, which corresponds to assortativity $r$ in the range $-0.18$ to 0.6. }
\label{lognormal}
\end{figure}

To demonstrate the impact of assortativity on the strong friendship paradox, we consider a network with $e(k,\kpr)$ that has a bivariate log-normal distribution with equal means $m$, equal variances $s^2$, and correlation coefficient $c$.
%In this case, $p(k)=k^{-2}\phi\left(\log k;m,s^2\right)e^{m-\frac{1}{2}s^2}$, where $\phi(\cdot)$ is the normal pdf.
The assortativity can be written as
\begin{equation}
r=\frac{\mathrm{Cov}(k,k')}{\mathrm{Var}(k)} =
\frac{e^{cs^2}-1}{e^{s^2}-1}.
%\mathrm{Cov}(k,k') = e^{2m+s^2}\left(e^{cs^2}-1\right)\hspace{0.4in}r=\frac{e^{cs^2}-1}{e^{s^2}-1}
\end{equation}
Note that the assortativity is bounded by $-e^{-s^2}\leq r\leq 1$, and
increases with $s$. We can then express $\mu_x(k)$ analytically as
\begin{align*}
\mu_x(k) &= P(k'>k|k)=P(\log k'>\log k|k)\nonumber \\
&= 1-\Phi\left\{\frac{\log k-E(\log k'|\log k)}{\sqrt{\mathrm{Var}(\log k'|\log k)}}\right\} \nonumber \\
&=1-\Phi\left\{\frac{(1-c)(\log k-m)}{\sqrt{(1-c^2)s^2}}\right\}\nonumber \\
&=1-\Phi\left\{\frac{\log k-m}{s}\sqrt{\frac{1-c}{1+c}}\right\}.
\end{align*}
% where $\Phi\{\cdot\}$ is the CDF of normal distribution.
% Since neighbors' degrees are independent in the 2K model, $\sigma_x^2(k)=\frac{1}{k}\mu_x(k)[1-\mu_x(k)]$.
It follows that $f(k)$ decreases with $k$.
%Figure~\ref{lognormal} shows how the paradox strength varies with assortativity.
As the network becomes more disassortative ($c<0$),
%the slope of $f(k)$ increases, with an increasingly sharp transition
$f(k)$ undergoes an increasingly sharp transition
from 1 to 0 around $k=e^m$ (Fig.~\ref{lognormal}(a)).  Given that most nodes
have low degree,
%this results in the global paradox being strongest when $r$ is lowest.
this leads to a globally stronger paradox
% \remove{for lower values of $r$}
in more disassortative networks (Fig.~\ref{lognormal}(b)), consistent with our prediction.
% We can see when the network is disassortative, then $\mu_x(k)$ increases if $k<e^m$, decreases if $k>e^m$
% \note{AP: I do not understand why the previous sentence is there. I
% suggest deleting it, unless there is a good reason. Misha: I think something is missing here. I wanted to write normally the network has larger amount of low degree nodes, this cause the global probability of paradox to increase.}
%\note{KL: this derivation does not use $f(k)$. Should we move it above Eq 5? Also, need to reference the figure and make the statement that negative assortativity magnifies paradox.}
%
% If the assumption of neighbors as i.i.d.~variables holds, from the central limit theorem, we know that
% For large $k$, the function $f(k)$ will be close to Gaussian, giving
% (see Figure~\ref{lognormal})
% \begin{align}
% \label{normalf}
% f(k)&=1-\Phi\left\{\frac{\frac{1}{2}-\mu_x(k)}{\sigma_x(k)}\right\}\\
% \label{normalf1}
% &=1-\Phi\left\{\frac{1-2\mu_x(k)}{2\sqrt{\mu_x(k)[1-\mu_x(k)]}}\sqrt{k}\right\}
% \end{align}
% Here we know that there is a scale of $\sqrt{k}$ in the parameter. In this formulation, if two different degree classes have the same mean $\mu(k_1)=\mu(k_2)$ and $k_1>k_2$, then this will turn $f(k_1)<f(k_2)$. However, in the real-world networks, we do not observe such phenomena in the probability of majority. \note{KL: I do not understand this argument.}

% Real-world networks
The structure of real-world networks creates conditions for the paradox. %nodes to observe a majority of neighbors with a higher degree.
We study six networks from a variety of domains, including social networks (friendship links on LiveJournal blogging site~\footnote{http://snap.stanford.edu/data/soc-LiveJournal1.html}, community structure on Youtube~\footnote{http://snap.stanford.edu/data/com-Youtube.html})
%(follow links on LiveJournal blogging site~\footnote{http://snap.stanford.edu/data/soc-LiveJournal1.html}, money lending relations on Prosper social finance site~\footnote{http://konect.uni-koblenz.de/networks/prosper-loans})
technological networks (Skitter internet graph~\footnote{http://snap.stanford.edu/data/as-skitter.html} and Google web hyperlink graph~\footnote{http://snap.stanford.edu/data/web-Google.html}), scientific citations graph (Arxiv~\footnote{http://snap.stanford.edu/data/cit-HepPh.html}), and relationships between English words~\footnote{http://wordnet.princeton.edu/}. These networks vary in size from 34.5K nodes (Arxiv) to almost 4M nodes (LiveJournal), and assortativity from 0.045 (LiveJournal) to -0.062 (English Words).
Table~\ref{tbl:realnets} reports the observed fraction of nodes in these networks who see a majority of their neighbors with a larger degree.  This fraction is very large in all networks, ranging from 75\% to 90\%.

\begin{table}[h]
\begin{tabular}{|l|l|c|c|c|}\hline
\textit{Network} & \textit{Type} & \textit{Observed} & \textit{3K model} & \textit{2K model}\\\hline\hline
LiveJournal & \small{Social} & 83.71\% & 84.43\% & 86.95\%\\\hline
Youtube & \small{Social} & 89.94\% & 88.51\% & 90.34\%\\\hline
Skitter & \small{Internet} & 88.62\% & 90.35\% & 95.79\%\\\hline
Google & \small{Web} & 77.31\% & 78.25\% & 84.36\%\\\hline
ArXiv HEP & \small{Citation} & 78.71\% & 79.67\% & 83.83\%\\\hline
English words & \small{Semantic} & 75.23\% & 71.00\% & 71.05\% \\\hline
%LiveJournal & \small{Social} & 83.71\% & 84.43\% & 85.22\%\\\hline
%Youtube & \small{Social} & 89.94\% & 88.51\% & 93.08\%\\\hline
%Skitter & \small{Internet} & 88.62\% & 90.35\% & 95.65\%\\\hline
%%Google & \small{Web} &   77.01\% & 78.08\% & 83.82\%\\\hline
%Google & \small{Web} & 77.31\% & 78.25\% & 85.32\%\\\hline
%%ArXiv HEP & \small{Citation} & 78.69\% & 79.64\% & 84.13\%\\\hline
%ArXiv HEP & \small{Citation} & 78.71\% & 79.67\% & 84.14\%\\\hline
%English words & \small{Semantic} & 75.23\% & 71.00\% & 76.58\% \\\hline
\end{tabular}
%%Prosper & \small{Finance} & 88.14\% & 88.26\% & 89.65\%\\\hline
%Prosper & \small{Finance} & 87.98\% & 88.12\% & 89.55\%\\\hline
\caption{Observed fraction of nodes in real-world networks that experience the strong friendship paradox, compared to predictions of the two proposed models.}
\label{tbl:realnets}
\end{table}

\begin{figure*}
\centering
\includegraphics[width=\textwidth]{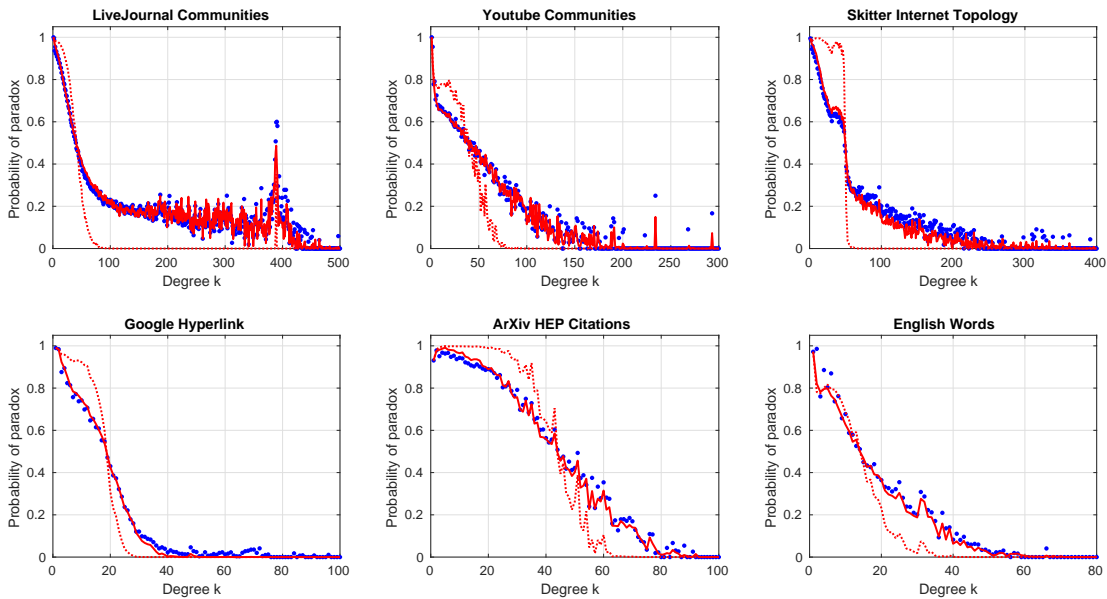}
\caption{(Color online) Probability of the strong friendship paradox in six real-world networks,
%Blue dots give the observed fraction of degree-$k$ nodes in the paradox regime. Dotted red line is the theoretical calculation given only degree-degree correlation.  Solid red line is the theoretical calculation given degree-degree and neighbor-neighbor correlations.
comparing observed fraction of degree-$k$ nodes that are in the paradox regime (blue dots) to predictions of the $2K$ model (dotted red line) and the $3K$ model (solid red line).
%The graphs from left to right are (1) LiveJournal: Friendships of users in \textit{LiveJournal} community, (2) Skitter: Data exchange between autonomous systems in the internet, (3) Google hyperlink: Web pages connected by hyperlinks, (4) Prosper Loans: Loans between users on \textit{Prosper.com}, the duplicated edges are merged into one, (5) ArXiv HEP: citation network on high energy physics papers on \textit{arXiv.org} (6) WordNet: English words in the dictionary linked by synonym, antonym, meronym, \textit{etc.} relations.
}
\label{paradox}
\end{figure*}

% While the
% binomial distribution (Eq.~\ref{f(k)-2K}) \remove{gives a fairly good estimate of}\note{overestimates} the
% global probability of the strong friendship paradox in a network
% predicted by the $2K$ model is close to its observed value
% (Table~\ref{tbl:realnets})
Table~\ref{tbl:realnets} shows that the observed fractions of nodes experiencing the paradox
are close to the global probabilities
predicted by the $2K$ model, when $\mu_x(k)$ is set to the actual  frequency with which a neighbor of a degree-$k$ node has larger degree.
However, a breakdown by degree class reveals 
significant deviations. Figure~\ref{paradox} plots the paradox
probability $f(k)$ for a degree-$k$ node (blue dots).
% KL - introduce critical degree
%The $2K$ estimate using Eq.~\ref{f(k)-2K} (dotted line) and the empirical value (dots) agree when the paradox probability is $0.5$. We define the degree at which this occurs as the critical degree $k_c$ of the network.
We define the degree at which the $2K$ estimate (Eq.~\ref{normalf}) of paradox probability is 0.5 as the critical degree $k_c$ of the network. By construction, $k_c=\mathrm{Median}(q(k))$.
%The $2K$ estimate using Eq.~\ref{f(k)-2K} (dotted line) overestimates the paradox for low degree nodes and underestimates it for higher degree nodes.
Nodes with degree $k<k_c$ are likely to experience the paradox, while those with $k>k_c$ are unlikely to do so.
The $2K$ model (dotted line) overestimates the paradox for low-degree nodes and underestimates it for high-degree nodes.
%Normally, since a network has many low-degree nodes, the $2K$ model overestimates the fraction of nodes in the paradox regime.
This suggests that %positive correlations between neighbor degrees are significant, and
the $2K$ model is insufficient, %.  We therefore need to
% This result indicates that the estimate of variance from the $2K$ structure is inaccurate, which also implies that the neighbors are in fact correlated. In this case, we need to
and we need to
take into account structure beyond degree correlations of connected pairs of nodes.
%to investigate one higher order structure beyond assortativity,
% to fix the value of variance of neighbors in each degree class.

%$3K$ structure has not been widely studied.

\paragraph*{3K Model.}
If neighbor degrees are identically distributed but \emph{correlated} random
variables, Eq.~(\ref{normalf}) %\ref{f(k)-2K})
must be modified to represent a
multivariate rather than a single binomial distribution.
To deal with the correlation, we now consider a pair of neighbors, with degree $k_i$ and $k_j$, of a single degree-$k$ node, and their indicator functions $x_i$ and $x_j$ as defined in Eq.~(\ref{ind}).
The corresponding multivariate normal approximation then gives
% approximation in Eq.~(\ref{normalf}) may be written more generally as
\[
f(k)=1-\Phi\left\{\frac{\frac{1}{2}-\mu_x(k)}{\sigma_x(k)}\right\},
\]
where the variance $\sigma_x^2(k)$
%, the variance of the (averaged) indicator
%function $\overline{x}$ for a degree-$k$ node,
is now
% \changed{Equation Changed!!}
\begin{align}
\label{3kvar}
\sigma_x^2(k) &=\mathrm{Var}(\bar{x}) = \mathrm{Var}\left(\frac{1}{k}\sum_{i=1}^kx_i\right)\nonumber\\
&=\frac{1}{k^2}\left[\sum_{i=1}^k\mathrm{Var}(x_i)+2\sum_{i=1}^{k-1}\sum_{j=i+1}^{k}\mathrm{Cov}(x_i,x_j)\right]\nonumber\\
% &=\frac{1}{k^2}\left[k\mathrm{Var}(x_i)+k(k-1)\;\mathrm{Cov}(x_i,x_j)\right]\nonumber\\
&=\frac{1}{k}\mu_x(k)[1-\mu_x(k)]+\frac{k-1}{k}\;\mathrm{Cov}(x_i,x_j).
%\frac{1}{n(k)}\sum_{\alpha=1}^{n(k)}(\bar{x}_\alpha-\bar{\bar{x}})^2\nonumber\\
%&=\frac{1}{k^2n(k)}\left[\sum_{\alpha=1}^{n(k)}\sum_{i=1}^k(x_{\alpha i}-\bar{\bar{x}})^2\right.\nonumber\\
%&\hspace{0.2in}+\frac{k-1}{k}\left.\sum_{\alpha=1}^{n(k)}\sum_{i\neq j}(x_{\alpha i}-\bar{\bar{x}})(x_{\alpha j}-\bar{\bar{x}})\right]\nonumber\\
%&=\frac{1}{k}\text{Var}(x_{\alpha i})+\frac{k-1}{k}\text{Cov}(x_{\alpha i},x_{\alpha j})
\end{align}

Unlike in Eq.~(\ref{normalf}), where $f(k)$ is completely
determined by $\mu_x(k)$, the $3K$ model requires
the covariance term to be specified.
Using values determined empirically from real-world networks as in the
2K model, we
obtain very accurate paradox probability estimates (solid line in
Figure~\ref{paradox}). These estimates also improve on the global 2K results shown in Table~\ref{tbl:realnets} for all cases except Youtube and English words, 
where the two estimates are nearly identical due to their close agreement for low degree values that represent a large fraction of nodes in the network.
%which %exhibit a significant 
%have large fractions of low-degree nodes where the normal approximation used in both 2K and 3K models is least accurate.

%To understand the effect of the covariance term,
%consider the joint degree distribution function %\note{of a connected triplet}
%$t(k^{\prime}_i,k,k^{\prime}_j)$ for finding a connected ordered triplet %, the probability of finding a connected
To understand the effect of the covariance term, consider the $3K$-distribution $t(k^{\prime}_i,k,k^{\prime}_j)$, the joint degree distribution of a connected ordered triplet of nodes with degrees $(k^{\prime}_i,k,k^{\prime}_j)$.
%, formed by a node of degree $k$ and its two neighbors with degrees $k^{\prime}_i$ and $k^{\prime}_j$:
Conditioning on the degree $k$ of the focal node gives the joint degree distribution of its two neighbors:
%\deleted{with degrees $k^{\prime}_i$ and $k^{\prime}_j$}:
% The network's $3K$-structure is captured by the joint distribution of degrees of three connected nodes, which can be written as $P(k_1,k,k_2)$. We condition the distribution on the central node of the triplet, which reduces to the degree correlations of neighbors of a degree $k$ node.
\begin{align}
P(k^{\prime}_i, k^{\prime}_j|k) &= P(k^{\prime}_i, k, k^{\prime}_j|k) = \frac{t(k^{\prime}_i, k, k^{\prime}_j)}{g(k)}\nonumber\\
g(k) &=\sum_{k^{\prime}_i, k^{\prime}_j}t(k^{\prime}_i, k, k^{\prime}_j).
\end{align}
The indicator function covariance term in Eq.~(\ref{3kvar}) is
\begin{align}
&\mathrm{Cov}(x_i,x_j)=P(x_i=1, x_j=1|k) - P(x_i=1|k)^2\nonumber\\
&\hspace{0.2in}= P(k^{\prime}_i>k, k^{\prime}_j>k|k) - P(k^{\prime}_i>k|k)^2,
\end{align}
where
\begin{equation}
P(k^{\prime}_i>k, k^{\prime}_j>k|k)=\frac{1}{g(k)}\sum_{k^{\prime}_i>k, k^{\prime}_j>k}t(k^{\prime}_i,k,k^{\prime}_j).
\end{equation}
and $P(k^{\prime}_i>k|k)$ is given by Eq.~(\ref{mu}).  Thus, the
covariance takes into account correlations only up to the level of
% The sum only counts
chains $(k^{\prime}_i,k,k^{\prime}_j)$.
Any higher-order correlations beyond $3K$, such as those
involving connected subgraphs of four nodes, %quadruplets of connected nodes,
would no longer be consistent
with a normal approximation for $f(k)$, since they would
involve information beyond the second moment of the indicator function.
The remarkable success of the $3K$ model in Fig.~\ref{paradox} suggests that
such higher-order correlations are not needed to explain the paradox,
or that they are negligible in real-world networks.

\begin{figure}[H]
\hspace{-0.3in}
\includegraphics[width=0.57\textwidth]{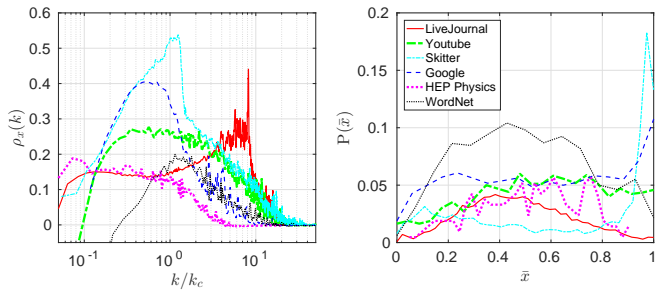}
\caption{(Color online) (Left) Neighbor-neighbor correlation coefficient by degree class for each network discussed in this paper. (Data have been smoothed). (Right) Distribution of $\bar{x}$ at the critical degree $k_c$.}
\label{correlation}
\end{figure}

Define the neighbor-neighbor correlation as
\begin{equation}
\rho_x(k)=\frac{\mathrm{Cov}(x_i,x_j)}{\sqrt{\mathrm{Var}(x_i)\mathrm{Var}(x_j)}}
=\frac{\mathrm{Cov}(x_i,x_j)}{\mu_x(k)[1-\mu_x(k)]}.
\end{equation}
Note that this correlation, like $\sigma_x(k)$, is based not on the neighbors' degrees but on the indicator function comparing them to the node's degree.
Figure~\ref{correlation} shows empirically determined values of
$\rho_x(k)$ for the real-world
networks we studied.  Recall that in the $2K$ model, the probability
that a degree-$k$ node has a neighbor with degree greater than $k$ is
determined completely by $e(k,k')$ and is unrelated to the degrees of
the other neighbors. %, and so $\rho_x(k)=0$.
%KL - 2K structure does not tell us anything about $\rho_x(k)$
One might reasonably expect
low-degree nodes to have mostly neighbors of higher degree, high-degree
nodes to have mostly neighbors of lower degree, and medium-degree nodes
to have a mix of both.  Figure~\ref{correlation}, however, depicts a
different scenario: medium-degree nodes prefer to have
neighbors with similar degree \emph{to one another}---whether those
neighbors have higher or lower degree.  To see how these correlations
may be indicative of
the macroscopic organization of a network, we plot the distribution of $\bar{x}$, the fraction of higher-degree neighbors, for %medium-degree
nodes with $k=k_c$. In the technological networks of Skitter and Google, such medium-degree nodes link more often to high-degree nodes, possibly reflecting a hierarchical
network structure with medium-degree at the top level and high-degree
nodes at the next level. The remaining networks show a broad
distribution of $\bar{x}$, %suggesting that medium-degree nodes link to both higher and lower-degree nodes. This is
consistent with a core-periphery network structure where medium-degree
nodes link to higher-degree nodes in the core and to lower-degree nodes in the periphery~\cite{rombach2014core,Newman2015core}.

% the neighbor-neighbor correlations for the index variable in the six real-world networks we studied. This quantity was obtained by solving Eq.~\ref{3kvar2} for $\rho_x(k)$. The trends suggest that nodes in middle degree classes either prefer to link with higher degree nodes or lower degree nodes, but not both. Such trends are coherent with the core-periphery model of the network~\cite{Porter}. The quantitative study shows that core scores are diverse for the nodes in middle degree classes. \note{Misha: Maybe local core?} These trends also attenuates the paradox in lower degree classes and amplifies it in middle degree classes.

%\subsubsection{Discussion}

The connection between local measurement bias and network structure 
revealed by the strong friendship paradox
is crucial for several reasons.
% Strong friendship paradox is not merely a curiosity, but may affect our understanding of network structure.
It is often impractical to observe large networks in their entirety: instead, researchers estimate network properties by exploring local neighborhoods of select nodes.
The paradox, however, may systematically bias local views of networks structure, including sampled degree distribution~\cite{Achlioptas06}.
The strong friendship paradox also affects measurements of information in networks. Consider a network where nodes have attributes and estimate their prevalence from local observations. When attribute and degree are correlated, the paradox can create an illusion that the attribute is common even when it is globally rare~\cite{Lerman16majority}.  Finally, quantifying %local bias of this kind 
measurement bias may be necessary for predicting the evolution of dynamic processes such as domain formation by majority rule in interacting spin systems~\cite{Krapivsky03}, or synchronization of frequencies in complex networks such as electrical power grids~\cite{Chertkov13}.  Accounting for neighbor-neighbor correlations could be instrumental to the success of network models for such systems.

In this paper, we have studied strong friendship paradox in networks, a phenomenon that distorts nodes' observations of local network structure. The paradox leads most nodes to observe that a majority of their neighbors have a larger degree than their own.
%Due to this paradox, many nodes in a network observe that a majority of their neighbors have a larger degree than their own degree.
%Strong friendship paradox can also bias nodes' observations of some trait, in cases when that trait is correlated with node degree. As an example, consider social networks, where high status people are likely to have more friends than the low status individuals. As a consequence of the paradox, many people will observe that most of their friends have higher status than they do, even where there are few high status people overall~\cite{Lerman16}.
%
%\begin{figure}[H]
%\centering
%\includegraphics[width=0.5\textwidth]{nncorr.eps}
%\caption{Neighbor-neighbor correlation coefficient by degree class for each network discussed in this paper. (Data have been smoothed)}
%\label{correlation}
%\end{figure}
%
We have developed an analytical model of the strong friendship paradox,
%We showed that in order to accurately estimate the strength of the paradox in a network, we need to account for higher order network structure, including node degree distribution ($1K$ structure), degree correlations of connected nodes ($2K$ structure), and neighbor-neighbor degree correlations ($3K$ structure).
enabling highly accurate predictions of its strength in networks. In contrast to Feld's friendship paradox~\cite{Feld91}, which exists in any network with variance in the degree distribution, %a skewed degree distribution,
the strong friendship paradox requires information about higher-order network structure. Specifically, negative correlations between degrees of connected nodes---given by network's $2K$ structure---will magnify the paradox, especially in networks with a skewed degree distribution. The impact of disassortativity, however, is modulated by degree correlations between nodes' neighbors. %, notably the network's $3K$ structure.
These correlations---given by network's $3K$ structure---are necessary to accurately quantify the paradox.
The success of the $3K$ model in explaining the paradox is consistent with the observation~\cite{Mahadevan06} that it is sufficient to capture known network properties.
In order to %quantify local bias and mitigate its impact on collective phenomena,
mitigate the effects of local measurement bias in networks,
it is important to account for
the strong friendship paradox and how it is impacted by
higher-order network structure.

%\begin{itemize}
%\item $3K$ information is required for estimating the strong friendship paradox.
%\item binomial model gives a fair estimation to the global probability of paradox. 3k model gives more accurate estimation on each class of degree.
%\item Discussion of $rho(k)=\frac{1}{k-1}\left\{k-\frac{\mu(k)[(1-\mu(k))]}{\sigma^2(k)}\right\}$ which increase with scale $k/<k>$ and goes from 0 to 1 by k
%\[\rho(k)=\frac{1}{k-1}\left\{\frac{k\sigma^2(k)}{\mu(k)[1-\mu(k)]}-1\right\}\]
%\end{itemize}

%\appendix
%\subsubsection{Dataset}
%\begin{table}[h]
%\begin{tabular}{|c|l|l|r|r|r|l|}\hline
%& \textit{Name} & \textit{Type} & \textit{Nodes} & \textit{Edges} & \textit{Assort.}\\\hline\hline
%1 & LiveJournal & Social & 3,997,962 & 34,681,189 & 0.045145\\\hline
%2 & Youtube & Social &  1,134,890 & 2,987,624 & -0.036910\\\hline
%3 & Skitter & Internet & 1,696,415 & 11,095,298 & $-0.081422$\\\hline
%4 & Google & Hyperlink & 875,713 & 4,322,051 & $-0.055089$\\\hline
%5 & ArXiv HEP & Citation & 34,546 & 420,877 & $-0.005943$\\\hline
%6 & WordNet & Semantic & 146,005 & 656,999 & $-0.006286$\\\hline
%\end{tabular}
%%4 & ProsperLoans & Finance & 89,269 & 3,394,979 & $-0.070122$\\\hline
%\caption{(1) LiveJournal Dataset (2) Skitter internet topology dataset (3) Google: web pages connected by hyperlinks, released in 2002 in Google Programming Contest. (4) Prosper Loans: loans between two accounts in \textit{prosper.com} ()}
%\end{table}

%\blindtext \cite{article-minimal}

\bibliographystyle{apsrev4-1} % Tell bibtex which bibliography style to use
\bibliography{references} % Tell bibtex which .bib file to use (this one is some example file in TexLive's file tree)

%\begin{thebibliography}{10}
%\bibitem{Feld91} S.~L. Feld, Am. J. of Sociol., \textbf{96}(6), 1464--1477 (1991)
%\bibitem{Baer91} J.~S. Baer, A.~Stacy, and M.~Larimer, J. of studies on alcohol, \textbf{52}(6), 580--586 (1991)
%
%\bibitem{Hodas13icwsm} N. Hodas, F. Kooti, and K. Lerman, Proc. 7th Int. AAAI Conf. on Weblogs And Social Media (2013)
%\bibitem{Kooti14icwsm} F. Kooti, N. Hodas, and K. Lerman, Intl. Conf. on Weblogs and Social Media (2014)
%
%\bibitem{Jo14} H.-H. Jo and Y.-H. Eom, Phys. Rev. E, \textbf{90} 022809 (2014)
%
%\bibitem{Lerman16} K. Lerman, X. Yan and X.-Z. Wu, PLOS ONE \textbf{0} 0000 (2016)
%
%\bibitem{Pinheiro14} F.~L. Pinheiro, M.~D. Santos, F.~C. Santos, J.~M. Pacheco, Phys. Rev. Lett., \textbf{112} 098702 (2014)
%
%\bibitem{Mahadevan06} P.~Mahadevan, D.~Krioukov, K.~Fall, A.~Vahdat, ACM SIGCOMM Comp. Comm. Rev., \textbf{36}(4) 135--146 (2006)
%\end{thebibliography}

\end{document}